\documentclass[pra,a4paper,amsmath,showkeys,preprintnumbers]{revtex4}
\usepackage{geometry}
\usepackage{bbm}
\usepackage{amssymb}
\usepackage{amsmath}
\usepackage{graphicx}
\usepackage{amsfonts}
\usepackage{hyperref}
\usepackage{oldgerm}
\usepackage{color}
\usepackage{epsfig}
\usepackage{here}

\newcommand{\bra}[1]{\ensuremath{\left\langle #1\right|}}
\newcommand{\ket}[1]{\ensuremath{\left|#1\right\rangle}}

\begin{document}
\title{Examining the dimensionality of genuine multipartite entanglement}
\author{Christoph Spengler$^1$}
\email{Christoph.Spengler@univie.ac.at}
\author{Marcus Huber$^{1,2}$}
\author{Andreas Gabriel$^1$}
\author{Beatrix C. Hiesmayr$^{1,3}$}
\affiliation{$^1$University of Vienna, Faculty of Physics, Boltzmanngasse 5, 1090 Vienna, Austria}
\affiliation{$^2$University of Bristol, Department of Mathematics, Bristol BS8 1TW, U.K.}
\affiliation{$^3$Masaryk University, Institute of Theoretical Physics and Astrophysics, Kotl\'a\v{r}sk\'a 2, 61137 Brno, Czech Republic}

\begin{abstract}
Entanglement in high-dimensional many-body systems plays an increasingly vital role in the foundations and applications of quantum physics. In the present paper, we introduce a theoretical concept which allows to categorize multipartite states by the number of degrees of freedom being entangled. In this regard, we derive computable and experimentally friendly criteria for arbitrary multipartite qudit systems that enable to examine in how many degrees of freedom a mixed state is genuine multipartite entangled.
\end{abstract}


\preprint{UWThPh-2011-19}

\maketitle

\section{Introduction}
Ever since its discovery more than seventy years ago, quantum entanglement has been considered as the central essence of quantum theory, forcing us to rethink our view of reality, locality and causality. It impressively highlights the non-local-realistic and contextual character of nature and thereby provides insights into the very foundations of physics. Moreover, during the last two decades, it has become more and more clear that entanglement can serve as a resource for future information processing technologies, such as quantum cryptography, dense coding, quantum teleportation and quantum computing. It is even argued that entanglement plays a role in quantum phase transitions \cite{Sachdev}, ionization processes \cite{Akoury}, high energy physics \cite{Hies} and light-harvesting complexes \cite{Sarovar}.

When it comes to studying quantum phenomena in diverse systems one is regularly confronted with the problems of how to detect, characterize and quantify entanglement. With the exception of bipartite qubit systems, these problems are in general extremely hard to solve for systems of arbitrary number of parties and dimensions, i.e. multipartite qudits.

In the present paper we focus on a finer characterization of \emph{genuine multipartite entanglement} \cite{Guehnedetection} in multilevel systems. Genuine multipartite entangled states have been shown to be vital for fundamental tests of quantum physics \cite{Mermin,GHZ,RelEntanglement} and find application in measurement-based quantum computing \cite{mbqc} and quantum secret sharing \cite{QSS,Schauer}. Although, this type of entanglement is not bounded on the dimensionality of the local systems, the use of systems with more than two levels, i.e. qudits, brings with it several advantages and deeper insights. For instance, it was found that quantum correlations are more robust against decoherence the more degrees of freedom are entangled \cite{Liu,CGLMP}. Qudit entanglement also improves the security of quantum key distribution \cite{CerfQKD}, and allows quantum secret sharing schemes \cite{Keet}, distributed protocols \cite{Fitzi,LiY} and error-correcting codes \cite{Looi} which cannot be realized with qubits. It is also to be expected that quantum computers that encode more than one qubit of information in each particle will require less resources and will thus be more efficient \cite{Zhou,Joo}. Furthermore, high-dimensional multipartite entanglement is essential for a complete understanding of quantum theory \cite{Bancal,Son,CerfGHZ,Lee,Barreiro}.

The aim of this paper is to provide practical and experimentally feasible criteria that allow to examine the dimensionality of genuine multipartite entanglement. The central problem is the following: Suppose we have realized a multipartite qudit scenario in the laboratory. How can we verify if a state is genuine multipartite entangled (GME) and how many degrees of freedom are involved in the entanglement?

For pure states this question is easily answered via the ranks of the reduced density matrices. However, for mixed states, as they appear in any real experiment, this is a nontrivial problem. Consider e.g. the mixed state $\rho_c=\frac{1}{2} \ket{GHZ_3}\bra{GHZ_3}+\frac{1}{6} \sum_{i=0}^{2}\ket{iii}\bra{iii}$, where $\ket{GHZ_3}=\frac{1}{\sqrt{3}}(\ket{000}+\ket{111}+\ket{222})$ is a tripartite Greenberger-Horne-Zeilinger (\emph{GHZ}) state entangled in three degrees of freedom. This state $\rho_c$ is not truly three-dimensionally entangled since it can be decomposed into $\rho_c=\frac{1}{3} (\ket{GHZ_{1,2}}\bra{GHZ_{1,2}}+\ket{GHZ_{1,3}}\bra{GHZ_{1,3}}+\ket{GHZ_{2,3}}\bra{GHZ_{2,3}})$ with $\ket{GHZ_{i,j}}=\frac{1}{\sqrt{2}}(\ket{iii}+\ket{jjj})$, which are each entangled in only two local degrees of freedom.

\section{Definitions}
Let us first give a precise definition of the dimensionality of multipartite entanglement -- a multipartite generalization of the Schmidt rank \cite{terhal,BrussSchmidt,BrussJMP} complementary to the tensor rank \cite{eisert} that allows to characterize multipartite qudit states. This generalization is important as the tensor rank is \emph{not} the crucial characteristic that the many-body entangled qudit states in \cite{Keet,Fitzi,LiY,Looi,Zhou,Joo,Bancal,Son,CerfGHZ,Lee,Barreiro} have in common. This is mainly because in the multipartite case
there is no simple connection between the tensor rank, $k$-separability and multilevel entanglement. For example, the states $\ket{W}=\frac{1}{\sqrt{3}}(\ket{001}+\ket{010}+\ket{100})$, $\ket{GHZ_3}=\frac{1}{\sqrt{3}}(\ket{000}+\ket{111}+\ket{222})$ and $\ket{\Psi_{\mbox{bisep.}}}=\ket{0}\otimes \frac{1}{\sqrt{3}}(\ket{00}+\ket{11}+\ket{22})$ all have tensor rank~$3$. However, only $\ket{GHZ_3}$ is commonly regarded as multilevel-multipartite entangled. Hence, for a finer categorization of multipartite entanglement it is needed to specify novel characteristic quantities.

For a pure $n$-partite state $\ket{\psi} \in \mathcal{H}=\mathbb{C}^{d_1}\otimes\ldots \otimes \mathbb{C}^{d_n}$ consider the set of all reduced density matrices $ \{ \rho_{A}=Tr_B(\ket{\psi}\bra{\psi}) \}$ regarding all bipartitions $ \gamma=\{(A|B)\}$. Evidently, iff the state $\ket{\psi}$ is fully separable then the rank of all reduced density matrices $\rho_{A}$ is $1$. On the other hand, the state $\ket{\psi}$ contains entanglement iff there exists a $\rho_{A}$ with $\mbox{rank}(\rho_A)>1$. For a fixed bipartition $(A|B)$, the dimensionality of entanglement is determined by the Schmidt rank \cite{terhal,BrussSchmidt,BrussJMP} which equals $\mbox{rank}(\rho_{A})$. Hence, iff $\max  \{\mbox{rank}(\rho_{A})\}=f$ with $f\geq2$ then the state $\ket{\psi}$ contains $f$-dimensional entanglement. We define a state to be $f$-dimensionally \emph{genuine multipartite entangled} (GME) iff it is at least $f$-dimensionally entangled with respect to \emph{all} bipartitions, that is $\min \{ \mbox{rank}(\rho_{A})\}=f$. This can be extended to mixed states in a natural way: A mixed state $\rho=\sum_i p_i \ket{\psi_i} \bra{\psi_i}$ is $f$-dimensionally entangled iff there exists no decomposition $\{p_i,\ket{\psi_i}\}$ into pure states $ \ket{\psi_i}$ of dimensionality $f_i$ all obeying $f_i<f$, but a decomposition into pure states satisfying $f_i \leq f$ for all $i$. A mixed state $\rho=\sum_i p_i \ket{\psi_i} \bra{\psi_i}$ is $f$-dimensionally GME iff any decomposition $\{p_i,\ket{\psi_i}\}$ of $\rho$ contains at least one GME state $\ket{\psi_i}$ of dimensionality $f$ or higher.

\section{Dimensionality criteria}
The problem of determining the dimensionality of multipartite entanglement for a given mixed state $\rho$ is as complex as the separability problem, since it is practically impossible to vary over all pure state decompositions of $\rho$. For this reason it is imperative to find computable criteria for the detection of high-dimensional genuine multipartite entanglement. First steps in this direction have recently been made by Lim \emph{et al.} \cite{Lim} and Li \emph{et al.} \cite{Li}. However, the criterion by Lim \emph{et al.} is only able to discriminate $2$-dimensional from $3$-dimensional genuine multipartite entanglement in tripartite three-level systems. The criterion by Li \emph{et al.} applies to arbitrary dimensionality and system size, but its noise resistance is rather unsatisfactory. Hence, further progress is needed here. Recently, a framework of criteria detecting genuine multipartite entanglement was introduced in \cite{hmgh,dicke,hsetal,ghrh}. Although it belongs to the strongest criteria for the detection of GME states without requiring semidefinite programming \cite{jungnitsch}, it does not discriminate states of different dimensionality. In the present paper, we show how this powerful framework can be extended for verifying the presence of high-dimensional genuine multipartite entanglement in arbitrary mixed states.

Consider a density matrix $\rho$ of an $n$-partite $d$-level system, i.e. a Hilbert space $\mathcal{H}=\left(\mathbb{C}^d \right)^{\otimes n}$. For a twofold copy $\rho^{\otimes 2}$ on $\mathcal{H}^{\otimes 2}$ we define for each bipartition $(A|B)$ of $\mathcal{H}$ a permutation operator $\mathcal{P}_{A}$ which permutes the subsystem $A$ with its copy $A'$, i.e.
\begin{align*}
\mathcal{H}^{\otimes 2}= \mathcal{H}_{A}\otimes\mathcal{H}_{B}\otimes\mathcal{H}_{A'}\otimes\mathcal{H}_{B'}\\
\stackrel{\mathcal{P}_{A}}{\longrightarrow} \mathcal{H}_{A'}\otimes\mathcal{H}_{B}\otimes\mathcal{H}_{A}\otimes\mathcal{H}_{B'}
\end{align*}
For instance, for the bipartition $(\{1\}|\{2,\ldots,n\})$ and the vector $\ket{k}^{\otimes n} \otimes \ket{l}^{\otimes n} \in \mathcal{H}^{\otimes 2}$ the corresponding operator $\mathcal{P}_{\{1\}}$ acts like
\begin{align*}
\mathcal{P}_{\{1\}} \ket{k}^{\otimes n} \otimes \ket{l}^{\otimes n} = \ket{l}\otimes\ket{k}^{\otimes (n-1)}\otimes\ket{k}\otimes\ket{l}^{\otimes (n-1)} \ .
\end{align*}
Using this abbreviation we introduce the quantity
\begin{align}
\label{Q0}
Q_0=\sum_{k \neq l}^{d-1}\left(|\bra{k_n}\rho\ket{l_n}|-\sum_{\gamma}\sqrt{\bra{k,l}\mathcal{P}_{A}^{\dagger}\rho^{\otimes2}\mathcal{P}_{A}\ket{k,l}}\right)
\end{align}
with $\ket{k_n}=\ket{k}^{\otimes n}$ and $\ket{k,l}=\ket{k}^{\otimes n} \otimes \ket{l}^{\otimes n}$, where the $\ket{k}, \ket{l}$ are vectors of an orthonormal basis $\{\ket{0},\ldots,\ket{d-1}\}$ of $\mathbb{C}^d$ and the sum runs over all bipartitions $\gamma=\{(A|B)\}$. Furthermore, we introduce the quantities $(m\in \{1,\ldots,\lfloor \frac{n}{2} \rfloor \})$
\begin{equation}
\label{Qm}
Q_m=\frac{1}{m}\left[\sum_{k,l=0}^{d-2}
\sum_{\sigma}\left(\underbrace{|\langle \alpha^{k}|\rho|\beta^{l}\rangle|}_{O^{k,l}_{\alpha,\beta}}-\sum_{\delta}\underbrace{\sqrt{\langle \alpha^{k}|\otimes \langle \beta^{l}|\mathcal{P}_\delta^{\dagger}\rho^{\otimes 2}\mathcal{P}_\delta|\alpha^{k} \rangle\otimes |\beta^{l}\rangle}}_{P^{k,l}_{\alpha,\beta}}\right)-N_D\sum_{l=0}^{d-2}\sum_{\alpha}\underbrace{\langle \alpha^{l}|\rho|\alpha^{l}\rangle}_{D^{l}_{\alpha}}\right]
\end{equation}
wherein $|\alpha^{l}\rangle \in \mathcal{H}$ are product vectors, where $m$ of the $n$ local systems contained in the set $ \alpha $ are in the state $\ket{l+1}$ and remaining ones in $\ket{l}$, i.e. $|\alpha|=m$ and
\begin{align}
|\alpha^{l}\rangle=\bigotimes_{i\in\alpha}\ket{l+1}_i\bigotimes_{i\notin\alpha}\ket{l}_i \ ,
\end{align}
and the same holds for $|\beta^{l}\rangle$. We have
\begin{align}
 \sigma&=\{(\alpha,\beta):|\alpha\cap\beta|=m-1\} \ , \\
 N_D&=(d-1) m (n-m-1) \ .
\end{align}
The innermost sum depends on $(\alpha,\beta)$ and runs over\footnote{Note that the empty set $\{\emptyset\}$ is always neglected.}
\begin{align}
\delta=\left\{\begin{array}{cl} \alpha & \hspace{0.5cm} \mbox{\ if \ }k=l \ , \\
\{\delta \ | \ \delta\subset\overline{\alpha \backslash \beta}\} & \hspace{0.5cm} \mbox{\ if \ } k<l \ , \\
\{\delta \ | \ \delta\subset\overline{\beta \backslash \alpha}\} & \hspace{0.5cm} \mbox{\ if \ } k>l \ , \end{array}\right.
\end{align}
where the complement (overline) is taken with respect to the set $\{1,\ldots,n\}$. Now, the main result of this paper is that if any of these functions $Q_m$ fulfills
\begin{align}
\label{criterion}
Q_m>f-2 \hspace{0.8cm} (\ \mbox{where} \ f \in  \{ 2,\ldots,d \} \ )
\end{align}
for a given density matrix $\rho$, then this state is at least $f$-dimensionally genuine multipartite entangled.

\section{Proof}
One standard strategy for detecting entanglement or distinguishing different types of entanglement is to introduce a quantity $Q(\rho)$ and to maximize it over all states of a specific type (e.g. $k$-separable states, states with particular Schmidt-rank, states of an entanglement class, etc.). Consequently, if for a particular state $\rho$ this maximum is exceeded, it necessarily must be of a different kind. The main problem in deriving entanglement criteria in this way is the involved maximization. The complexity of this problem can be reduced by using quantities $Q(\rho)$ which are convex in $\rho$, because in this case the optimization has to be performed over all pure states only. Nevertheless, the difficulty of finding the global maximum remains.

In the present paper, a completely new approach is used. Namely, the convex quantities $Q_m$ are constructed by incorporating the matrix elements of specific $f$-dimensionally GME states. This is done in a way such that by construction any other state of same or lower dimensionality cannot reach a certain bound. Thus, to prove that $Q_m \leq f-1$ holds for all states which are entangled in equal or less than $f$ degrees, no maximization has to be carried out.

First, consider the $f$-dimensionally genuine $n$-partite entangled \emph{GHZ} state
\begin{align}
\ket{GHZ_f}=\frac{1}{\sqrt{f}}\sum_{i=0}^{f-1}\ket{i}^{\otimes n} \ . \label{eq_ghzstate}
\end{align}
In density matrix form $\rho_{fGHZ}=\ket{GHZ_f}\bra{GHZ_f}$, the only nonzero elements are $\bra{k}^{\otimes n}\rho_{fGHZ}\ket{l}^{\otimes n}=\frac{1}{f}$. Each term $|\bra{k_n}\rho\ket{l_n}|$ in (\ref{Q0}) singles out the absolute value of an off-diagonal element of $\rho_{fGHZ}$, such that
\begin{align}
\sum_{k \neq l}|\bra{k_n}\rho_{fGHZ}\ket{l_n}|=2 {f\choose 2} \frac{1}{f} =f-1 \ .
\end{align}
As can easily be confirmed, all terms $\sqrt{\bra{k,l}\mathcal{P}_{A}^{\dagger}\rho_{fGHZ}^{\otimes2}\mathcal{P}_{A}\ket{k,l}}$ vanish for any choice of $k,l$ and any bipartition $(A|B)$ as the corresponding matrix elements are all zero. Thus, we have shown that $Q_0=f-1$ for $\rho_{fGHZ}$. In addition, it was proven in \cite{hmgh} that
\begin{align}
\label{remark1}
|\bra{k_n}\rho\ket{l_n}|-\sum_{\gamma}\sqrt{\bra{k,l}\mathcal{P}_{A}^{\dagger}\rho^{\otimes2}\mathcal{P}_{A}\ket{k,l}} \leq 0 \ ,
\end{align}
holds for all biseparable states. Hence, each of these terms (\ref{remark1}) can only be larger than zero if the state $\rho$ is genuine multipartite entangled in $\ket{k}$ and $\ket{l}$. Now, since in (\ref{Q0}) the absolute values of all off-diagonal elements of $\rho_{fGHZ}$ are added up, and since all terms which are subtracted from this sum are zero for $\rho_{fGHZ}$, it follows that $\ket{GHZ_f}$ is the only $f$-dimensionally GME pure state that reaches $Q_0=f-1$. Thus, any state that exceeds $f-1$ must at least be $(f+1)$-dimensionally GME, which proves (\ref{criterion}) for $m=0$.

\begin{figure}[t]
\label{temperature}
\includegraphics[width=12cm]{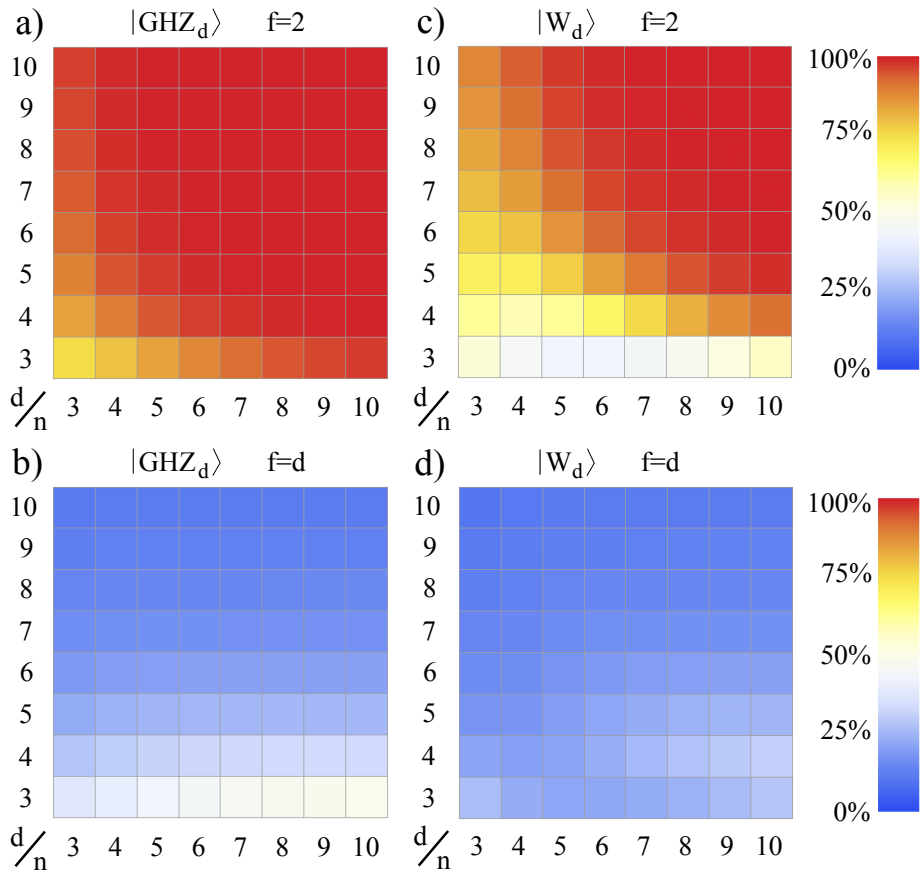}
\setlength{\unitlength}{1cm}%
\caption{(Color online) The noise resistance for the states $\ket{GHZ_d}$ and $\ket{W_d}$. The scale indicates up to which value of white noise a state is detected to be $f$-dimensionally genuine multipartite entangled (GME). a) up to the illustrated thresholds for $p$, the state $\rho=\frac{p}{d^n}\mathbbm{1}+(1-p)\ket{GHZ_d}\bra{GHZ_d}$ is detected to be GME, i.e. $Q_0>0$. b) for noise $p$ below the illustrated thresholds the state $\rho$ is detected by $Q_0>d-2$ to be truly $d$-dimensionally GME. c) and d) illustrate these thresholds for the state $\rho=\frac{p}{d^n}\mathbbm{1}+(1-p)\ket{W_d}\bra{W_d}$ using $Q_1(\rho)>f-2$.}
     \label{temperature}
\end{figure}

Due to the way it is constructed, the function (\ref{Q0}) is optimally suited to detect high-dimensional genuine multipartite entanglement in mixed states which are close to $GHZ$ states. To show that the quantities $Q_m$ serve their purpose (\ref{criterion}) for $m>0$, we introduce the $f$-dimensionally genuine multipartite entangled $m$-\emph{Dicke state} $( m\in \{1,\ldots,\lfloor \frac{n}{2} \rfloor \} )$
\begin{align}
|D^m_f\rangle:=\frac{1}{\sqrt{(f-1){n\choose m}}}\sum_{l=0}^{f-2}\sum_{\alpha}\bigotimes_{i\in\alpha}\ket{l+1}_i\bigotimes_{i\notin\alpha}\ket{l}_i \ ,
\end{align}
where the inner sum runs over all $\alpha$ with $|\alpha|=m$ (see also \cite{dicke}). Note that this includes a generalization $\ket{W_f}=|D^1_f\rangle$ of the prominent $W$ state for qudits\footnote{Note that a different generalization of the $W$ state was introduced in \cite{Sanders}. However, the state introduced therein is local-unitarily equivalent to the original $W$ state, and therefore does not contain higher-dimensional entanglement.}. First, observe that for any fixed choice of $k=l$ in $Q_m$, (\ref{Qm}) reduces to the inequalities from Refs.~\cite{hmgh,dicke}, i.e. in this case it is proven that $Q_m \leq 0$ holds for all biseparable states. On the other hand, by summing over all $k$ and $l$ in $Q_m$ we add up the absolute value of specific off-diagonal elements $O^{k,l}_{\alpha,\beta}$ of the $m$-\emph{Dicke state} $|D_f^m\rangle$.
From these off-diagonal elements (determined by the proper set $\sigma$) there are corresponding diagonal elements (labeled $P^{k,l}_{\alpha,\beta}$) subtracted, which correspond to biseparable states having the same off-diagonal elements. For a subset of all those off-diagonal elements this suffices (as with the inequality based on $Q_0$), however, for some there are no corresponding diagonal elements belonging to a biseparable state. In order to guarantee that $Q_m\leq0$ for all biseparable states one also needs to subtract the corresponding diagonal elements of the \emph{Dicke state} (labeled $D^l_{\alpha}$). Counting the cardinality of this subset is a purely combinatorial problem (similar to \cite{dicke}) resulting in the factor $N_D$. By construction, this guarantees for fixed dimensionality, the maximal value of $Q_m$ for the corresponding \emph{Dicke state}, as in this case the sum of the off-diagonal elements is maximal, whereas all $P^{k,l}_{\alpha,\beta}$ are zero. By scaling this maximum with the constant $\frac{1}{m}$ we can unify all quantities $Q_m$ in one consistent framework, i.e. the only $f$-dimensionally GME pure state that can attain $Q_m=f-1$ is $|D^m_f\rangle$.

\begin{figure}[t]
\centering
\includegraphics[width=12cm]{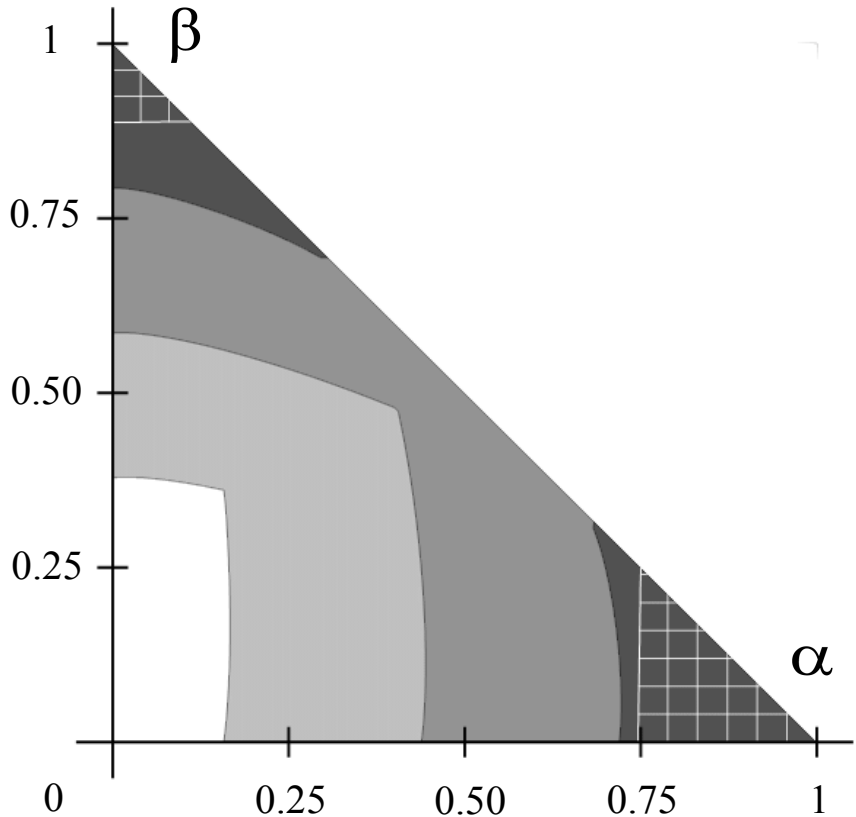}
\caption{Illustration of the detection strength of $Q_0, Q_1>f-2$ for the tripartite four-level state $\rho = \alpha \ket{GHZ_4}\bra{GHZ_4} + \beta \ket{W_4}\bra{W_4} + \frac{1-\alpha-\beta}{64} \mathbbm{1}$. The dark gray region is detected to be $4$-dimensionally GME (In comparison, fidelity-based criteria merely detect the white meshed part of this region to be $4$-dimensionally GME). The middle gray region is detected to be at least $3$-dimensionally GME, and the light gray region is detected to be at least $2$-dimensionally GME.}\label{fig2}
\end{figure}

\section{Detection strength}
The introduced criteria allow to examine the dimensionality of multipartite entanglement in a noisy environment. Fig.~\ref{temperature} shows the robustness for the states $\ket{GHZ_d}$ and $\ket{W_d}$ in the presence of white noise. We compared the illustrated thresholds with the thresholds of entanglement witnesses that follow from the fidelity of a state (see \cite{Li,BrussSchmidt}), i.e. a state is $d$-dimensionally GME if $\bra{GHZ_d} \rho \ket{GHZ_d} > \frac{d-1}{d}$ or $\bra{W_d} \rho \ket{W_d} > \frac{n(d-1)-1}{n(d-1)}$, respectively. Here, we found that our criteria are strictly stronger for all $d>2$ and $n>2$ -- specifically, they outperform the noise robustness of previously known criteria \cite{Li} for \emph{GHZ}-like states. E.g., the tripartite state
\begin{align}
\rho=(1-p)\ket{GHZ_3}\bra{GHZ_3}+ p \frac{1}{27}\mathbbm{1} \ ,
\end{align}
is detected to be $3$-dimensionally GME by $Q_0>1$ for $0\leq p<0.375$, whereas $\bra{GHZ_3} \rho\ket{GHZ_3}>\frac{2}{3}$ merely detects the range $0\leq p<0.346$. For
\begin{align}
\rho=(1-p)\ket{W_3}\bra{W_3}+ p \frac{1}{27}\mathbbm{1} \ ,
 \end{align}
 the difference is even more significant: Using $Q_1>1$, we detect the range $0\leq p<0.265$ in comparison to $0\leq p<0.173$ following from $\bra{W_3} \rho\ket{W_3}>\frac{5}{6}$. A further example of the detection strength is given in Fig.~\ref{fig2}. As can be seen therein, the region of states detected by our criteria is considerably larger than the region revealed by fidelity-based witnesses. Finally, let us stress that the quantities $Q_m$ are by construction optimally suited to detect $f$-dimensional GME states which are close to \emph{GHZ}, \emph{W} and \emph{Dicke states}. If instead an unclassified input state is given one can improve the detection by maximizing the outcome of $Q_m$ over local-unitary transformations. An appropriate optimization scheme can be found in \cite{spengler1,spengler2,spengler3}.

\section{Conclusion}
Creating high-dimensional multipartite entangled states is one of the current challenges in experiments on quantum physics. In the present paper, we gave a precise mathematical characterization of such states and provided criteria for the dimensionality of genuine multipartite entanglement applicable to arbitrary multi-qudit systems. These criteria are easily computable since they do not rely on semidefinite programming or eigenvalue computations, but only on functions of density matrix elements. They are also advantageous in experiments, as they are rather robust against noise and to apply them it is not necessary to determine the entire density matrix of the system under consideration. In detail, due to the fact that the quantities $Q_m$ only involve the matrix elements of \emph{GHZ}, \emph{W} and \emph{Dicke states}, it is merely needed to determine these few entries of the density matrix, which can always be achieved via local measurements and corresponding correlations (see also the discussion in e.g. \cite{hmgh,dicke,hsetal,ghrh}). Consequently, they can be experimentally implemented with a reduced number of local observables, since the number of measurements for a full quantum state tomography scales exponentially in the number of parties $n$, i.e. is of the order $\mathcal{O}( d^{2n})$ \cite{thew}, whereas the number of density matrix elements that occur in $Q_m$ is only of the order $\mathcal{O}(d^2{n\choose m})$, that is polynomial in $n$ (Note that the notation in terms of two-fold copies of a state is only a matter of compactness, i.e. in experiments it is not necessarily needed to have two copies at a time). Finally, it is noteworthy that our results are even promising to be closely related to measures of genuine multipartite entanglement, as e.g. for multipartite qubits the quantity $Q_0$ yields a strong lower bound on the \emph{gme-concurrence} \cite{gmeconc}.

\begin{acknowledgements}
C.S. and M.H. acknowledge financial support from the Austrian Science Fund (FWF) -- Project P21947N16. A.G. was supported by a research grant of the University of Vienna.
\end{acknowledgements}

\end{document}